\documentclass[aps,preprint,superscriptaddress,amsfonts,amsmath,a4paper,showpacs]{revtex4}
\usepackage{graphicx}
\usepackage{color}
\begin{document}

\title{Cluster Glasses of Semiflexible Ring Polymers}

\author{Mohammed Zakaria Slimani}
\affiliation{Donostia International Physics Center, Paseo Manuel de Lardizabal 4,
E-20018 San Sebasti\'{a}n, Spain.}

\author{Petra Bacova}
\affiliation{Departamento de F\'{\i}sica de Materiales, Universidad del Pa\'{\i}s Vasco (UPV/EHU),
Apartado 1072, E-20080 San Sebasti\'{a}n, Spain.}
\affiliation{Materials Physics Center MPC, Paseo Manuel de Lardizabal 5, E-20018 San Sebasti\'{a}n, Spain.}

\author{Marco Bernabei}
\affiliation{Donostia International Physics Center, Paseo Manuel de Lardizabal 4,
E-20018 San Sebasti\'{a}n, Spain.}
\affiliation{Departament de Fisica Fonamental, Universitat de Barcelona,
Mart\'{\i} i Franqu\`{e}s 1, E-08028 Barcelona, Spain.}

\author{Arturo Narros}
\affiliation{Faculty of Physics, University of Vienna, Boltzmanngasse 5, A-1090 Vienna, Austria.}

\author{Christos N. Likos}
\affiliation{Faculty of Physics, University of Vienna, Boltzmanngasse 5, A-1090 Vienna, Austria.}

\author{Angel J. Moreno}
\email[Corresponding author: ]{wabmosea@ehu.es}
\affiliation{Donostia International Physics Center, Paseo Manuel de Lardizabal 4,
E-20018 San Sebasti\'{a}n, Spain.}
\affiliation{Materials Physics Center MPC, Paseo Manuel de Lardizabal 5, E-20018 San Sebasti\'{a}n, Spain.}
\affiliation{Centro de F\'{\i}sica de Materiales (CSIC, UPV/EHU), Paseo Manuel de Lardizabal 5, E-20018 San Sebasti\'{a}n, Spain.}

\begin{abstract}

We present computer simulations of concentrated solutions of unknotted nonconcatenated
semiflexible ring polymers. Unlike in their flexible counterparts,
shrinking involves a strong energetic penalty, favoring interpenetration and clustering 
of the rings. We investigate the slow dynamics of the centers-of-mass of the rings in the amorphous cluster phase,
consisting of disordered columns of oblate rings penetrated by bundles of prolate ones.  
Scattering functions reveal a striking decoupling of self- and collective motions. Correlations between
centers-of-mass exhibit slow relaxation, as expected for an incipient glass
transition, indicating the dynamic arrest of the cluster positions. However,
self-correlations decay at much shorter time scales. This feature is a manifestation of the
fast, continuous exchange and diffusion of the individual rings over the matrix of clusters. 
Our results reveal a novel scenario of glass-formation in a simple monodisperse system,
characterized by self-collective decoupling, soft caging and mild dynamic heterogeneity.

\end{abstract}
\date{\today}
\maketitle

Over the last years, the fascinating properties of ring polymers have attracted the interest  
of  researches in broad disciplines of physics, chemistry, biophysics and 
mathematics \cite{bielawski:science:2002,dobay:pnas:2003,grosberg:pnas:2004,hirayama:jpa:2009,vettorel:pb:2009,marenduzzo:jpcm:2010,milner:prl:2010,micheletti:physrep:2011}.
The simple operation of joining permanently the two ends of a linear chain, forming a
ring, has a dramatic impact in its structural and dynamic properties.
This includes differences with linear chains in, e.g., their swelling \cite{jang:jcp:2003}, 
rheological \cite{kapnistos:nm:2008} or scaling behavior \cite{halverson:jcp:2011a}.
Another remarkable effect of the ring topology is the non-Gaussian character
of the effective potential in solution \cite{narros:sm:2010,bohn:jcp:2010}, in contrast to
the well-known Gaussian potential found for linear chains \cite{louis:prl:2000}.

The use of effective potentials reduces real macromolecular solutions to effective fluids of ultrasoft,
fully-penetrable particles \cite{louis:prl:2000,likos:pr:2001,gottwald:prl:2004,mladek:prldendrim}.
This methodology facilitates the investigation of the physical properties of polymers in solution.
The investigation of tunable {\it generic} models of ultrasoft particles,
inspired by the bounded character of the real effective interactions in polymer solutions,
offers a route for discovering and designing novel soft matter phases with potential
realizations in real life. For a family of generic models, the so-called $Q^{\pm}$-class \cite{likos:jcp:2007,mladek:prl:2006}, in which 
the Fourier transform of the  bounded potential is non positive-definite, the ultrasoft
particles can form clusters. At sufficiently high densities
the fluid transforms into a cluster crystal \cite{likos:jcp:2007,mladek:prl:2006}.
However, the approach based on effective potentials derived {\it at infinite dilution}
has severe limitations at high concentrations, due to the emergence of many-body forces arising, e.g.,
from particle deformations. This has been recently demonstrated for the case of flexible ring polymers \cite{narros:sm:2010}.

%

In recent work, some of us have extended the study of Ref.~\cite{narros:sm:2010}
to the case of {\it semiflexible} rings \cite{bernabei:sm:2013}. 
Unlike in flexible rings, the presence of intramolecular
barriers  makes shrinkage energetically unfavourable. If semiflexible rings are sufficiently small, their size
is only weakly perturbed  \cite{bernabei:sm:2013}.  
This may facilitate interpenetration and promote clustering in order to fill the space in dense solutions.
This was not the case for very small rings due to excluded-volume effects, or for sufficiently long ones
in which the expected random arrangement of the centers-of-mass was recovered.
However, in a certain range of molecular weight an amorphous cluster phase was found,
consisting of disordered columns of oblate rings penetrated by bundles of prolate rings 
(see Figs.~12 and 13 in Ref.~\cite{bernabei:sm:2013}). 
This novel cluster phase emerges in a real, {\it one-component}, polymer solution with {\it purely repulsive} 
interactions \cite{bernabei:sm:2013}. This finding is crucially different from 
other soft matter cluster phases where clustering is mediated by short-range attraction
and long-range repulsion  \cite{cardinaux:jpcb:2011}. 
Although clustering of the rings was predicted by the obtained effective potential,
the anisotropic character of the real clusters was not captured by the {\it isotropic} effective
interaction, which did not incorporate the relative orientation between rings
as an additional, relevant degree of freedom \cite{bernabei:sm:2013}.

Recent simulations of a polydisperse (preventing crystallization)
generic fluid  of  ultrasoft, purely repulsive particles
of the $Q^{\pm}$-class, have revealed the possibility of forming a {\it cluster glass} \cite{coslovich:jcp:2012}. 
Whether this dynamic scenario may find a realization
in a real polymer solution is an open question. Apart from the eventual inaccuracy
of the ultrasoft potentials to describe real structural correlations at high concentrations
(see above), predictions on the dynamics can be misleading. Even by using the correct
mean-force potential describing  exactly the static correlations, the real dynamics
can be strongly influenced by the so-called transient forces \cite{briels:sm:2009},  related to the removed intramolecular
degrees of freedom and not captured by the mean-force potential.

Motivated by the emergence of the anisotropic cluster state in dense solutions of semiflexible rings,
in this Letter we investigate the associated dynamics in this phase.
We find a striking decoupling of self- and collective motions.
As expected for an incipient glass transition, correlations between centers-of-mass  exhibit slow relaxation, 
reflecting the dynamic arrest of the cluster positions.
However, self-correlations relax at much shorter time scales. 
This feature is a manifestation of the
fast, continuous exchange and diffusion of the individual rings over the quasi-static matrix of clusters.
Our results reveal a novel dynamic scenario for glass formation in a real,
simple monodisperse system,
characterized by the simultaneous presence of self-collective decoupling, soft caging and
mild dynamic heterogeneity.

We simulate  $N_{\rm R}=1600$ unknotted  nonconcatenated rings of $N=50$ monomers by using the 
Kremer-Grest model \cite{kremergrest:jcp:1990} with added bending stiffness. 
Model and simulation details are extensively described 
in Ref.~\cite{bernabei:sm:2013} (here we use a friction $\gamma =2$ instead of
$\gamma = 0.5$ used in \cite{bernabei:sm:2013}).
To estimate the characteristic ratio $C_\infty$ \cite{Rubinstein:Book2003}, we simulate the linear counterparts without 
excluded volume interactions between non-connected beads. 
In these conditions the chains show the expected Gaussian scaling at long contour distances, and
we obtain   $C_\infty$ from the long-$|i-j|$ plateau in $\langle R^2 (i-j) \rangle/|i-j|$,
with $R(i-j)$ the distance between monomers $i$ and $j$. We find $C_{\infty} \sim   10$, a value
typical of common stiff polymers \cite{Rubinstein:Book2003}. 
By simple scaling, we expect similar trends for biopolymers with $C_{\infty}\sim 1000 $
and polymerization degree  $N \sim 5000$.

By focusing on the structure and dynamics of the centers-of-mass of the rings, we use the average diameter of gyration
at infinite dilution, $D_{\rm g0}$, to normalize the density of the ring solution. 
Thus, we define the density as $\rho = N_{\rm R}(L/D_{\rm g0})^{-3}$, 
with $L$ the simulation box length. For $N=50$ we find $D_{\rm g0} = 13\sigma$, with $\sigma=1$ 
the monomer size \cite{bernabei:sm:2013}.
We explored a concentration range from  $\rho \rightarrow 0$  to  $\rho =20$.
The value  $\rho =20$ corresponds to a monomer density of $\rho_{\rm m} = 0.45 $, 
about half  the melt density in similar bead-spring models \cite{kremergrest:jcp:1990}.

\begin{figure}
\includegraphics[width=0.75\linewidth]{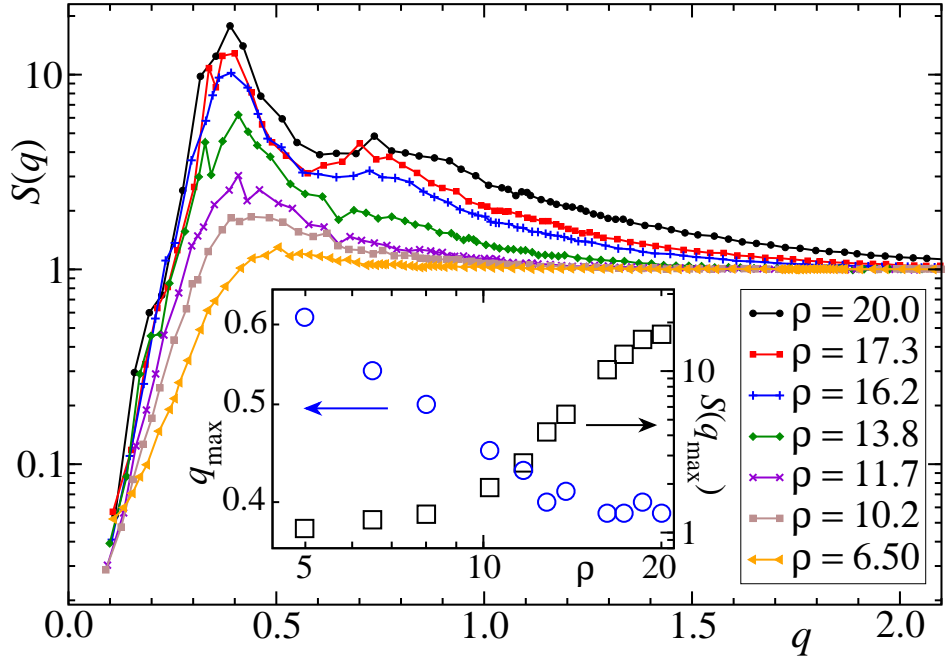} 
\newline
\caption{
Static structure factor $S(q)$ of the centers-of-mass (main panel), for different densities (see legend).
The inset shows the density dependence of $q_{\rm max}$ (circles) and $S(q_{\rm max})$ (squares),
where $q_{\rm max}$ is the wavector at the maximum of $S(q)$. 
Both $q_{\rm max}$ and $S(q_{\rm max})$ are estimated by fitting the main peak to a Gaussian. 
The corresponding error bars are smaller than the symbol sizes in the inset.}
\label{fig:sq}
\end{figure}

Fig.~S1 in the Supplemental Material shows results for the radial distribution 
function $g(R)$ of the centers-of-mass of the rings,
at different densities. Clustering at high densities is evidenced by the increasing maximum of $g(R)$ at zero distance.
Fig.~\ref{fig:sq} shows results for the static structure factor of the centers-of-mass,
$S(q)=N_{\rm R}^{-1}\langle  \sum_{j,k} \exp[i{\bf q}\cdot({\bf R}_j(0)-{\bf R}_k(0))]\rangle$, 
with ${\bf R}_{j,k}$ denoting positions of the centers-of-mass.
By increasing the concentration, $S(q)$ develops a sharp maximum at wavevector $q_{\rm max} \sim 0.4$.
This corresponds to a typical distance between centers-of-mass of $d \sim 2\pi/q_{\rm max} \sim 16$.
This is slightly higher than the typical diameter of gyration 
in the whole investigated density range ($12.4 < D_{\rm g} < 13.6$) \cite{bernabei:sm:2013}.
In simple liquids the main peak is followed by a pronounced minimum $S(q_{\rm min}) <1$ 
and higher-order harmonics \cite{hansen_theory_2006}.
Instead, we find a nearly featureless, smoothly decaying shoulder 
extending up to large $q$-values. This reflects the full interpenetrability of the rings at short distances.
The inset of Fig.~\ref{fig:sq} shows the peak height $S(q_{\rm max})$ (squares) versus the density.
The slope of $S(q_{\rm max})$ exhibits an abrupt change at $\rho\sim 10$.
We identify this feature as the onset of the cluster phase.
The maximum of $S(q)$ exhibits remarkable features.
Thus, it reaches values of up to $S(q_{\rm max}) \sim 20$ at the highest investigated densities. However,
these are not accompanied by crystallization, as would be expected by the Hansen-Verlet criterion for
simple liquids \cite{hansenverlet:pr:1969}.
Although the effective potential does not fully capture all details of the cluster structure 
(in particular its anisotropic character \cite{bernabei:sm:2013}),
Fig.~\ref{fig:sq} reveals a key feature of cluster-forming fluids
of fully-penetrable objects \cite{mladek:prl:2006,likos:jcp:2007}.
Namely, the wavevector  $q_{\rm max} \sim 0.4$ for the maximum of $S(q)$ (circles in the inset)
is esentially density-independent in the cluster phase. Thus, adding rings to the system
does not modify the distance between clusters ($d \sim 2\pi/q_{\rm max}$) but just their population \cite{mladek:prl:2006,likos:jcp:2007}.

\begin{figure}
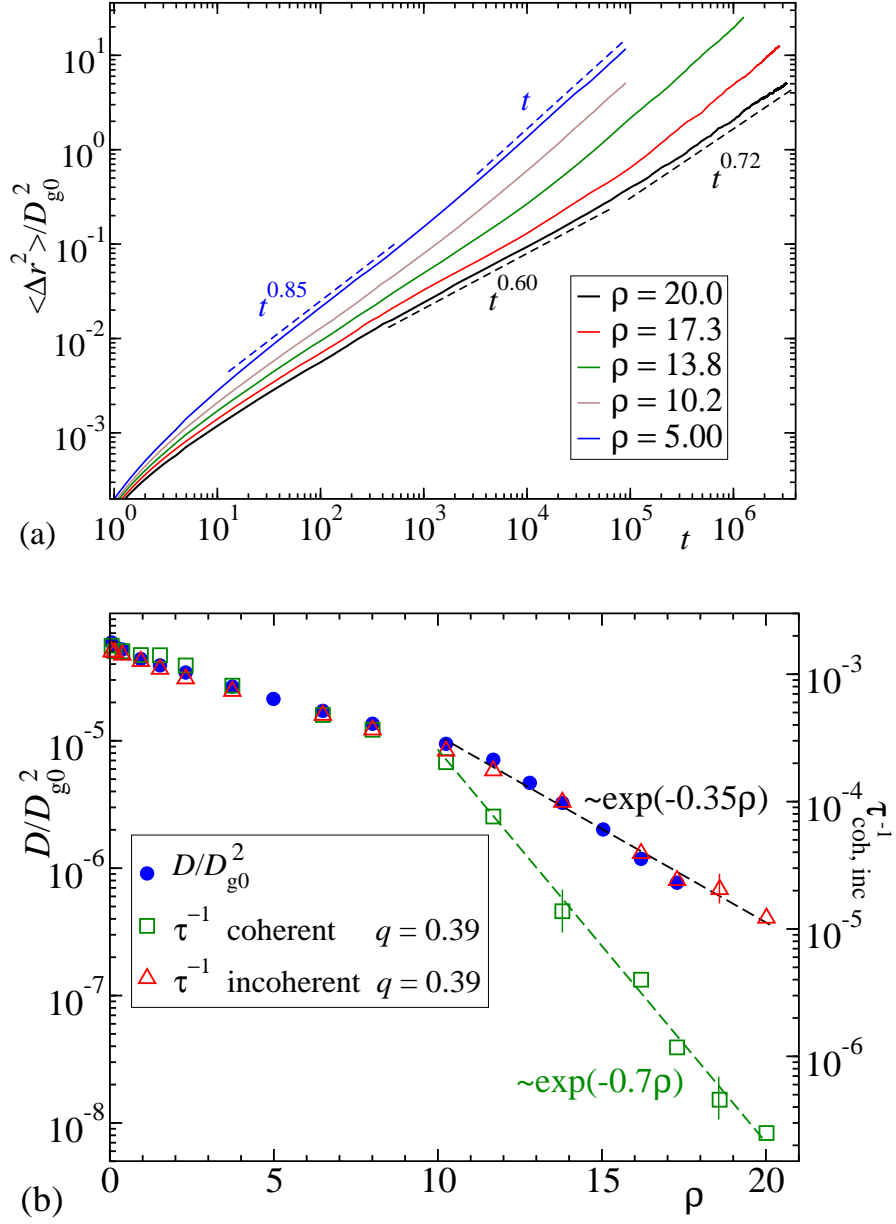

\begin{center}
\hspace{5 mm}\includegraphics[width=0.63\linewidth]{Fig2a.eps} 
\newline
\newline
\includegraphics[width=0.70\linewidth]{Fig2b.eps} 
\end{center}
\caption{(a): MSD of the centers-of-mass (solid lines), normalized by $D^2_{\rm g0}$,
for different densities (see legend).  Dashed lines describe approximate power-law behavior $\sim t^x$ 
(exponents given in the panel).
(b): Density dependence of the diffusivity, $D$,  and inverse relaxation times, $\tau^{-1}$. Some typical error bars are given.
Closed circles: $D$ normalized by $D^2_{\rm g0}$. Open symbols: $\tau^{-1}$ for the coherent (squares) 
and incoherent (triangles) scattering functions at $q = 0.39$. Left and right ordinate axes correspond to data of $D/D^2_{\rm g0}$ and $\tau^{-1}$, respectively. Both ordinate axes span over a same factor $2 \times 10^4$
for a fair comparison between different data sets. 
The dashed lines indicate apparent exponential dependence  $D, \tau^{-1} \sim \exp(-\Gamma\rho)$. 
Values of $\Gamma$ are given in the panel.}
\label{fig:msd}
\end{figure}

Now we investigate the slow dynamics of the rings in the cluster phase.
In standard molecular and colloidal fluids close to a glass 
transition \cite{binder_glassy_2005}, 
particles can be mutually trapped by their neighbors  over several time decades.
This is the well-known caging effect, which leads to a plateau in the mean squared displacement
(MSD, $\langle \Delta r^2 \rangle$) versus time $t$. The temporal extent of the caging regime increases on
approaching the glass transition (usually by increasing density and/or decreasing temperature).
At longer times, particles escape from the cage and reach the diffusive regime $\langle \Delta r^2 \rangle \propto t$.
Fig.~\ref{fig:msd}a shows the MSD 
of the centers-of-mass at different densities up to the highest investigated one.
Data are normalized by $D^2_{\rm g0}$ in order to show displacements in terms of the typical ring size.
In all cases, displacements at the end of the simulation correspond to several times the ring size.
Within the investigated concentration range, no plateau is found in the MSD.
A soft caging effect is observed, which is manifested as an apparent subdiffusive 
regime $\langle \Delta r^2 \rangle \sim t^x$, with $x<1$ decreasing by increasing concentration.
The crossover to diffusive behavior is found, in most cases, when displacements approach the typical ring 
size, $\langle \Delta r^2 \rangle \lesssim D^2_{\rm g0}$. However, this is not the case for the highest
investigated density $\rho = 20$, where a crossover to an apparent second subdiffusive regime is found,
persisting at least up to values of $\langle \Delta r^2 \rangle = 5 D^2_{\rm g0}$. The eventual
crossover to diffusion is beyond the simulation time scale.

Fig.~\ref{fig:msd}b shows the density-dependence of the diffusivity, $D$, of the centers-of-mass of the rings. 
This is determined 
as the long-time limit of $\langle \Delta r^2 \rangle/6t$, for the densities at which the 
linear regime $\langle \Delta r^2 \rangle \propto t$ is reached within the simulation time scale. 
A sharp dynamic crossover is found at $\rho \sim 10$, i.e., around the
density for the onset of the cluster phase (Fig.~\ref{fig:sq}). This crossover is characterized by
a much stronger density-dependence of the diffusivity in the cluster phase ($\rho > 10$) 
and, as we discuss below, a decoupling of self- and collective motions.
In the investigated density range of the cluster phase, we find an apparent exponential law
$D \sim \exp(-0.35\rho)$, which may suggest activated dynamics. Still, this conclusion
must be taken with care because of the limited range of observation (one decade in diffusivity).

\begin{figure}
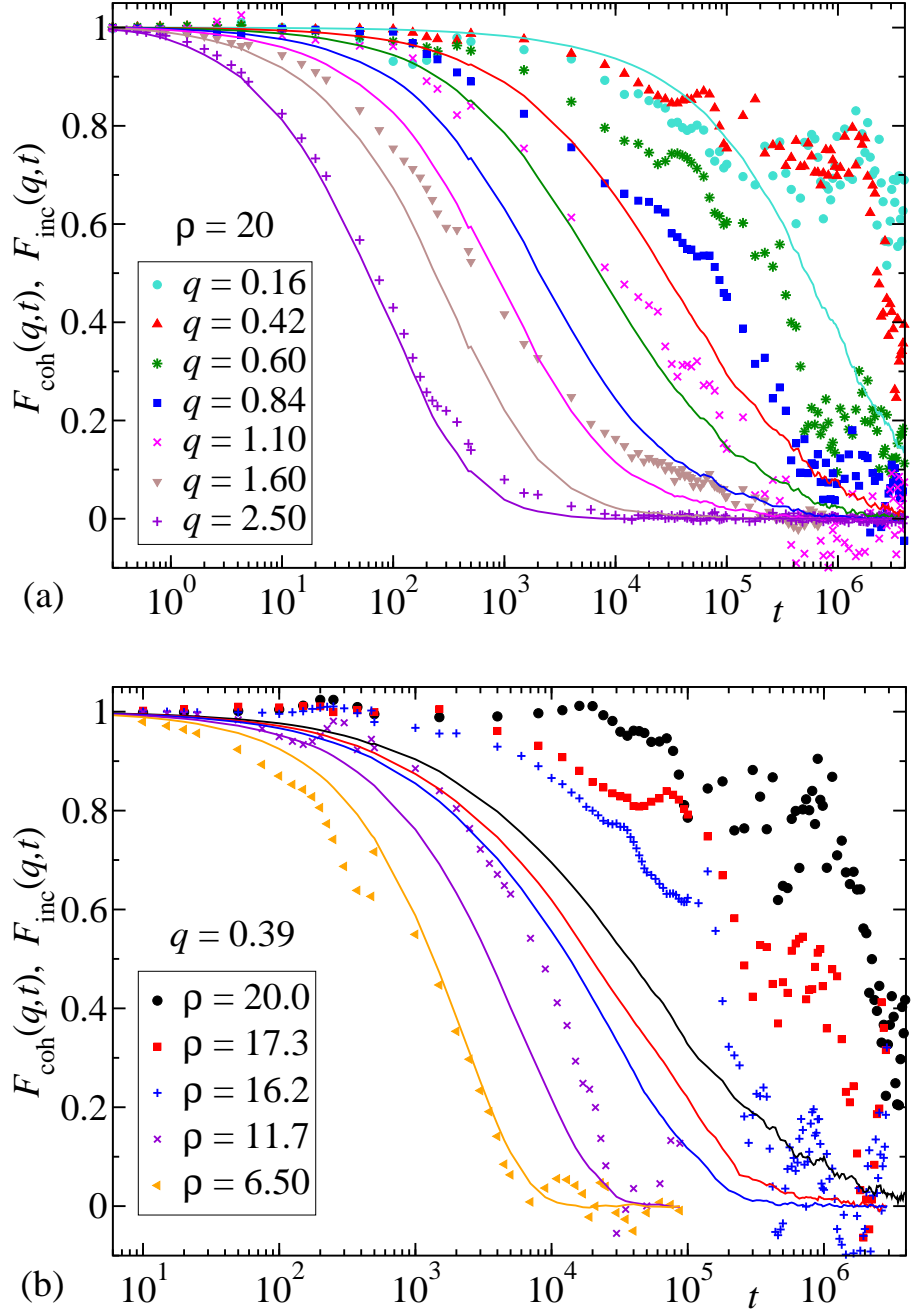

\includegraphics[width=0.72\linewidth]{Fig3a.eps} 
\newline
\newline
\includegraphics[width=0.72\linewidth]{Fig3b.eps} 
\newline
\caption{Scattering functions for the centers-of-mass of the rings. Symbols and lines correspond 
to coherent and incoherent functions, respectively. 
(a): Results for the highest investigated $\rho = 20$, and different
$q$-values. (b): Results for fixed $q = 0.39 \approx q_{\rm max}$ and different densities.
In each panel, two data sets with identical colors correspond to the coherent (symbols) 
and incoherent (line) function for a same value
of $q$ (in panel (a)) or $\rho$ (in panel (b)); see legends.}
\label{fig:fqt}
\end{figure}

Further insight on the dynamics can be obtained by computing scattering functions of the centers-of-mass.
Normalized coherent and incoherent functions are defined as $F_{\rm coh}(q,t) = [N_{\rm R}S(q)]^{-1}\langle  \sum_{j,k} \exp[i{\bf q}\cdot({\bf R}_j(t)-{\bf R}_k(0))]\rangle$ and $F_{\rm inc}(q,t) = N^{-1}_{\rm R}\langle  \sum_{j} \exp[i{\bf q}\cdot({\bf R}_j(t)-{\bf R}_j(0))]\rangle$, respectively.
Coherent functions probe pair correlations between centers-of-mass of the rings, whereas incoherent functions
probe self-correlations. Fig.~\ref{fig:fqt}a shows results for both functions at 
the highest investigated density $\rho=20$ and for
several representative wavevectors. Comparison between data sets reveals an unusual  result:
the incoherent functions relax in much shorter time scales than their coherent counterparts.
Only in the limit of large wavevectors $q \gg q_{\rm max}$, where no collective correlations
are really probed, both functions trivially approach each other.
We illustrate this effect by representing, for $\rho =20$, the $q$-dependence of the relaxation times $\tau$
of the scattering functions (see Fig.~S2 in the Supplemental Material). These are defined as the times
for which $F_{\rm coh, \rm inc} (q,\tau) = e^{-1}$.
Fig.~\ref{fig:fqt}b shows, for fixed wavevector $q = 0.39 \approx q_{\rm max}$, coherent and incoherent
scattering functions at several densities. In Fig.~\ref{fig:msd}b we show the density dependence 
of the respective inverse relaxation times, $\tau_{\rm coh,inc}^{-1}$. 
As can be seen, the time scale separation between coherent and incoherent functions
is associated to the onset of the cluster phase at $\rho \sim 10$, and becomes more pronounced by increasing the density. Within the whole investigated range, the incoherent inverse 
relaxation times follow the same density dependence as the diffusivity (note that both ordinate axes in Fig.~\ref{fig:msd}b span over a same factor $2 \times 10^4$ for a fair comparison between different data sets). In the cluster phase the inverse coherent times follow a much stronger dependence, with an apparent activation energy of about twice that of the diffusivity and incoherent inverse time.

Fig.~\ref{fig:fqt} demonstrates that
collective correlations slow down  by increasing density, 
reflecting the arrest of the cluster positions. This is the signature of an incipient glass transition.
However, unlike in simple glass-formers,
this is not accompanied by a similar arrest of the self-motions, which exhibit a much faster relaxation.
This reflects that fast, continuous exchange and diffusion of the rings takes place over the slowly
relaxing matrix of clusters. This is consistent with the  soft character of the caging regime
in the MSD (Fig.~\ref{fig:msd}a). 
As discussed in Ref.~\cite{bernabei:sm:2013},
clusters are not formed in the limit of small and large rings.
In Fig.~S3 of the Supplemental Material we show results for $g(R)$ and $S(q)$
in the former two limits of non-cluster forming rings (highest investigated densities 
for $N= 20$ and 100 in Ref.~\cite{bernabei:sm:2013}).
Fig.~S4 of the Supplemental Material shows the corresponding scattering functions.
No decoupling is observed there. This further supports the intimate relation between 
the formation of the cluster phase and the decoupling of self- and collective motions.
The small differences between coherent and incoherent functions in the non-cluster forming
systems can be roughly understood by  simple de Gennes narrowing \cite{hansen_theory_2006}, $\tau_{\rm coh}/\tau_{\rm inc} \sim S(q)$.
This is clearly not the case in the cluster phase (see Fig.~S5 in the Supplemental Material), confirming the highly non-trivial nature of the 
observed decoupling.

The dynamic scenario observed for {\it real} semiflexible rings, in the cluster phase, exhibits strong similarities
with results in cluster glass-forming fluids of {\it generic} fully-penetrable ultrasoft particles \cite{coslovich:jcp:2012}. 
These include the  crossover in the diffusivity to apparent activated behavior and the
decoupling between coherent and incoherent dynamics in the cluster phase.
Interestingly, the scenario observed for the semiflexible rings also has analogies with the dynamics
in two-component systems with very strong dynamic asymmetry \cite{horbach:prl:2002,moreno_relaxation_2006,mayer:macro:2009}, 
and more generally in crowded environments \cite{crowded-jpcm},
even if clustering and penetrable (`ultrasoft') character may be absent 
in such systems \cite{horbach:prl:2002,moreno_relaxation_2006}. 
Subdiffusive regimes in the MSD of the fast particles are usually observed in such mixtures,
extending up to distances much larger than the particle size. The trend in Fig.~\ref{fig:msd}a
for $\rho =20$ resembles this feature. Decoupling of self- and collective dynamics 
in the mentioned mixtures
is found for the fast component (`tracer'). The tracers perform large-scale fast diffusion along 
paths spanning over the confining matrix (formed by the slow component). 
Because of the slowly relaxing character of the matrix and the paths, 
collective correlations between the tracers decay in a much slower fashion than the self-correlations
\cite{horbach:prl:2002,moreno_relaxation_2006,mayer:macro:2009}. 

The results presented here for cluster-forming semiflexible rings constitute a {\it novel} realization of this decoupling scenario. First, it takes place in a real {\it monodisperse} system. This feature is intimately
connected to the fully penetrable character of the rings, which can behave both
as fast `tracers' moving from one cluster to other, and as part of the slow `matrix' formed by the cluster structure.
Second, it is not connected to the presence of strong dynamic heterogeneities, unlike
in the mentioned dynamically asymmetric mixtures \cite{horbach:prl:2002,moreno_relaxation_2006,mayer:macro:2009} where a clear distinction between `fast' and `slow' particles exists.
One might still think of a small fraction of rings performing much faster dynamics than the average, 
as a sort of `defect diffusion'. If this were the case the van Hove self-correlation function $G_{\rm s}(r,t)$ 
of the centers-of-mass would show, at long times, 
a strongly localized sharp main peak (owing to the majority slow rings), plus a secondary unlocalized peak or a broad tail corresponding to the minority fraction of fast rings. Fig.~S6 in the Supplemental Material displays
$G_{\rm s}(r,t)$ (symbols) for $\rho = 20$. 
This shows a smooth evolution with time. For comparison we include the results for simple Gaussian functions (lines)
with the same values of $\langle \Delta r^2  (t)\rangle$. Even in the most non-Gaussian case ($t = 10^6$),
no putative division into two subpopulations of minority `fast' and majority `slow' rings can be made.
This is further corroborated by the fact that the diffusivity and the inverse incoherent time
feature the same density dependence (Fig.~\ref{fig:msd}b). This is not the case
in systems with strong dynamic heterogeneity, in which diffusivities and relaxation times
are dominated by fast and slow particles, respectively.

As shown in Ref.~\cite{bernabei:sm:2013}, the cluster phase is formed by two subpopulations
of rings with very different shape.
The clusters consist of disordered columns of oblate rings (prolateness parameter $p \rightarrow -1$) penetrated
by bundles of elongated, prolate rings ($p \rightarrow 1$). It might still be argued that the initial prolateness of the ring
play a significant role in its ulterior (fast or slow) dynamics. We find that this is not the case either.
We have divided the rings into different sets
according to their $p$-values at $t=0$. Fig.~S7 in the Supplemental Material displays the MSD,
at $\rho = 20$, for several sets covering the whole $p$-range.
Very weak differences are observed  between the different sets. The most prolate rings are somewhat faster at early times,
suggesting some enhanced longitudinal motion of the elongated bundles. However all sets collapse for  
displacements smaller than the ring size.
In summary, the former results indicate that all rings participate in a similar fashion, via continuous 
exchange between clusters, in the relaxation of the self-correlations, without any clear distinction between
fast and slow subpopulations. This fast  mechanism 
weakly alters the cluster structure, which relaxes at much longer time scales,
leading to incoherent-coherent decoupling.

Although special techniques for the synthesis of pure rings have been developed \cite{bielawski:science:2002},
the usual, high-throughput approaches inadvertently result
into the presence of residual linear chains  \cite{kapnistos:nm:2008}.
Having noted this, the qualitative picture observed here for the dynamics of the pure rings will not be affected.
We performed additional simulations of a symmetric mixture of rings and linear counterparts
of identical $N=50$ (results will be presented elsewhere).
Though for identical total densities less pronounced effects are observed, we anticipate that
the rings in the mixture exhibit all the qualitative trends of Figs.~\ref{fig:sq}-\ref{fig:fqt}.

In summary, we have characterized slow dynamics in the amorphous cluster phase of 
a concentrated solution of unknotted nonconcatenated semiflexible rings.  
Our results reveal a novel dynamic scenario for glass formation in a real, simple monodisperse system,
characterized by the simultaneous presence of self- and collective decoupling, soft caging and
mild dynamic heterogeneity.

This work has been supported by the Austrian Science Fund (FWF), Grant No. 23400-N16.
We acknowledge generous allocation of CPU time
in CESCA (Spain).

\vspace{-2 mm}

\begin{thebibliography}{32}
\expandafter\ifx\csname natexlab\endcsname\relax\def\natexlab#1{#1}\fi
\expandafter\ifx\csname bibnamefont\endcsname\relax
  \def\bibnamefont#1{#1}\fi
\expandafter\ifx\csname bibfnamefont\endcsname\relax
  \def\bibfnamefont#1{#1}\fi
\expandafter\ifx\csname citenamefont\endcsname\relax
  \def\citenamefont#1{#1}\fi
\expandafter\ifx\csname url\endcsname\relax
  \def\url#1{\texttt{#1}}\fi
\expandafter\ifx\csname urlprefix\endcsname\relax\def\urlprefix{URL }\fi
\providecommand{\bibinfo}[2]{#2}
\providecommand{\eprint}[2][]{\url{#2}}

\bibitem[{\citenamefont{Bielawski et~al.}(2002)\citenamefont{Bielawski,
  Benitez, and Grubbs}}]{bielawski:science:2002}
\bibinfo{author}{\bibfnamefont{C.~W.} \bibnamefont{Bielawski}},
  \bibinfo{author}{\bibfnamefont{D.}~\bibnamefont{Benitez}}, \bibnamefont{and}
  \bibinfo{author}{\bibfnamefont{R.~H.} \bibnamefont{Grubbs}},
  \bibinfo{journal}{Science} \textbf{\bibinfo{volume}{297}},
  \bibinfo{pages}{2041} (\bibinfo{year}{2002}).

\bibitem[{\citenamefont{Dobay et~al.}(2003)\citenamefont{Dobay, Dubochet,
  Millett, Sottas, and Stasiak}}]{dobay:pnas:2003}
\bibinfo{author}{\bibfnamefont{A.}~\bibnamefont{Dobay}},
  \bibinfo{author}{\bibfnamefont{J.}~\bibnamefont{Dubochet}},
  \bibinfo{author}{\bibfnamefont{K.}~\bibnamefont{Millett}},
  \bibinfo{author}{\bibfnamefont{P.-E.} \bibnamefont{Sottas}},
  \bibnamefont{and} \bibinfo{author}{\bibfnamefont{A.}~\bibnamefont{Stasiak}},
  \bibinfo{journal}{Proc.~Natl.~Acad.~Sci.~U.S.A.}
  \textbf{\bibinfo{volume}{100}}, \bibinfo{pages}{5611} (\bibinfo{year}{2003}).

\bibitem[{\citenamefont{Moore et~al.}(2004)\citenamefont{Moore, Lua, and
  Grosberg}}]{grosberg:pnas:2004}
\bibinfo{author}{\bibfnamefont{N.~T.} \bibnamefont{Moore}},
  \bibinfo{author}{\bibfnamefont{R.~C.} \bibnamefont{Lua}}, \bibnamefont{and}
  \bibinfo{author}{\bibfnamefont{A.~Y.} \bibnamefont{Grosberg}},
  \bibinfo{journal}{Proc.~Natl.~Acad.~Sci.~U.S.A.}
  \textbf{\bibinfo{volume}{101}}, \bibinfo{pages}{13431}
  (\bibinfo{year}{2004}).

\bibitem[{\citenamefont{Hirayama et~al.}(2009)\citenamefont{Hirayama,
  Tsurusaki, and Deguchi}}]{hirayama:jpa:2009}
\bibinfo{author}{\bibfnamefont{N.}~\bibnamefont{Hirayama}},
  \bibinfo{author}{\bibfnamefont{K.}~\bibnamefont{Tsurusaki}},
  \bibnamefont{and} \bibinfo{author}{\bibfnamefont{T.}~\bibnamefont{Deguchi}},
  \bibinfo{journal}{J.~Phys.~A: Math.~Theor.} \textbf{\bibinfo{volume}{42}},
  \bibinfo{pages}{105001} (\bibinfo{year}{2009}).

\bibitem[{\citenamefont{Vettorel et~al.}(2009)\citenamefont{Vettorel, Grosberg,
  and Kremer}}]{vettorel:pb:2009}
\bibinfo{author}{\bibfnamefont{T.}~\bibnamefont{Vettorel}},
  \bibinfo{author}{\bibfnamefont{A.~Y.} \bibnamefont{Grosberg}},
  \bibnamefont{and} \bibinfo{author}{\bibfnamefont{K.}~\bibnamefont{Kremer}},
  \bibinfo{journal}{Phys. Biol.} \textbf{\bibinfo{volume}{6}},
  \bibinfo{pages}{025013} (\bibinfo{year}{2009}).

\bibitem[{\citenamefont{Marenduzzo et~al.}(2010)\citenamefont{Marenduzzo,
  Micheletti, and Orlandini}}]{marenduzzo:jpcm:2010}
\bibinfo{author}{\bibfnamefont{D.}~\bibnamefont{Marenduzzo}},
  \bibinfo{author}{\bibfnamefont{C.}~\bibnamefont{Micheletti}},
  \bibnamefont{and}
  \bibinfo{author}{\bibfnamefont{E.}~\bibnamefont{Orlandini}},
  \bibinfo{journal}{J.~Phys.: Condens.~Matter} \textbf{\bibinfo{volume}{22}},
  \bibinfo{pages}{283102} (\bibinfo{year}{2010}).

\bibitem[{\citenamefont{Milner and Newhall}(2010)}]{milner:prl:2010}
\bibinfo{author}{\bibfnamefont{S.~T.} \bibnamefont{Milner}} \bibnamefont{and}
  \bibinfo{author}{\bibfnamefont{J.~D.} \bibnamefont{Newhall}},
  \bibinfo{journal}{Phys. Rev. Lett.} \textbf{\bibinfo{volume}{105}},
  \bibinfo{pages}{208302} (\bibinfo{year}{2010}).

\bibitem[{\citenamefont{Micheletti et~al.}(2011)\citenamefont{Micheletti,
  Marenduzzo, and Orlandini}}]{micheletti:physrep:2011}
\bibinfo{author}{\bibfnamefont{C.}~\bibnamefont{Micheletti}},
  \bibinfo{author}{\bibfnamefont{D.}~\bibnamefont{Marenduzzo}},
  \bibnamefont{and}
  \bibinfo{author}{\bibfnamefont{E.}~\bibnamefont{Orlandini}},
  \bibinfo{journal}{Phys.~Rep.} \textbf{\bibinfo{volume}{504}},
  \bibinfo{pages}{1} (\bibinfo{year}{2011}).

\bibitem[{\citenamefont{Jang et~al.}(2003)\citenamefont{Jang, {{\c
  C}a${\breve{{\rm g}}}$in}, and {Goddard III}}}]{jang:jcp:2003}
\bibinfo{author}{\bibfnamefont{S.~S.} \bibnamefont{Jang}},
  \bibinfo{author}{\bibfnamefont{T.}~\bibnamefont{{{\c C}a${\breve{{\rm
  g}}}$in}}}, \bibnamefont{and} \bibinfo{author}{\bibfnamefont{W.~A.}
  \bibnamefont{{Goddard III}}}, \bibinfo{journal}{J.~Chem.~Phys.}
  \textbf{\bibinfo{volume}{119}}, \bibinfo{pages}{1843} (\bibinfo{year}{2003}).

\bibitem[{\citenamefont{Kapnistos et~al.}(2008)\citenamefont{Kapnistos, Lang,
  Vlassopoulos, Pyckhout-Hintzen, Richter, Cho, Chang, and
  Rubinstein}}]{kapnistos:nm:2008}
\bibinfo{author}{\bibfnamefont{M.}~\bibnamefont{Kapnistos}},
  \bibinfo{author}{\bibfnamefont{M.}~\bibnamefont{Lang}},
  \bibinfo{author}{\bibfnamefont{D.}~\bibnamefont{Vlassopoulos}},
  \bibinfo{author}{\bibfnamefont{W.}~\bibnamefont{Pyckhout-Hintzen}},
  \bibinfo{author}{\bibfnamefont{D.}~\bibnamefont{Richter}},
  \bibinfo{author}{\bibfnamefont{D.}~\bibnamefont{Cho}},
  \bibinfo{author}{\bibfnamefont{T.}~\bibnamefont{Chang}}, \bibnamefont{and}
  \bibinfo{author}{\bibfnamefont{M.}~\bibnamefont{Rubinstein}},
  \bibinfo{journal}{Nat.~Mater.} \textbf{\bibinfo{volume}{7}},
  \bibinfo{pages}{997} (\bibinfo{year}{2008}).

\bibitem[{\citenamefont{Halverson et~al.}(2011)\citenamefont{Halverson, Lee,
  Grest, Grosberg, and Kremer}}]{halverson:jcp:2011a}
\bibinfo{author}{\bibfnamefont{J.~D.} \bibnamefont{Halverson}},
  \bibinfo{author}{\bibfnamefont{W.~B.} \bibnamefont{Lee}},
  \bibinfo{author}{\bibfnamefont{G.~S.} \bibnamefont{Grest}},
  \bibinfo{author}{\bibfnamefont{A.~Y.} \bibnamefont{Grosberg}},
  \bibnamefont{and} \bibinfo{author}{\bibfnamefont{K.}~\bibnamefont{Kremer}},
  \bibinfo{journal}{J. Chem. Phys.} \textbf{\bibinfo{volume}{134}},
  \bibinfo{pages}{204904} (\bibinfo{year}{2011}).

\bibitem[{\citenamefont{Narros et~al.}(2010)\citenamefont{Narros, Moreno, and
  Likos}}]{narros:sm:2010}
\bibinfo{author}{\bibfnamefont{A.}~\bibnamefont{Narros}},
  \bibinfo{author}{\bibfnamefont{A.~J.} \bibnamefont{Moreno}},
  \bibnamefont{and} \bibinfo{author}{\bibfnamefont{C.~N.} \bibnamefont{Likos}},
  \bibinfo{journal}{Soft Matter} \textbf{\bibinfo{volume}{6}},
  \bibinfo{pages}{2435} (\bibinfo{year}{2010}).

\bibitem[{\citenamefont{Bohn and Heermann}(2010)}]{bohn:jcp:2010}
\bibinfo{author}{\bibfnamefont{M.}~\bibnamefont{Bohn}} \bibnamefont{and}
  \bibinfo{author}{\bibfnamefont{D.~W.} \bibnamefont{Heermann}},
  \bibinfo{journal}{J.~Chem.~Phys.} \textbf{\bibinfo{volume}{132}},
  \bibinfo{pages}{044904} (\bibinfo{year}{2010}).

\bibitem[{\citenamefont{Louis et~al.}(2000)\citenamefont{Louis, Bolhuis,
  Hansen, and Meijer}}]{louis:prl:2000}
\bibinfo{author}{\bibfnamefont{A.~A.} \bibnamefont{Louis}},
  \bibinfo{author}{\bibfnamefont{P.~G.} \bibnamefont{Bolhuis}},
  \bibinfo{author}{\bibfnamefont{J.~P.} \bibnamefont{Hansen}},
  \bibnamefont{and} \bibinfo{author}{\bibfnamefont{E.~J.}
  \bibnamefont{Meijer}}, \bibinfo{journal}{Phys.~Rev.~Lett.}
  \textbf{\bibinfo{volume}{85}}, \bibinfo{pages}{2522} (\bibinfo{year}{2000}).

\bibitem[{\citenamefont{Likos}(2001)}]{likos:pr:2001}
\bibinfo{author}{\bibfnamefont{C.~N.} \bibnamefont{Likos}},
  \bibinfo{journal}{Phys.~Rep.} \textbf{\bibinfo{volume}{348}},
  \bibinfo{pages}{267} (\bibinfo{year}{2001}).

\bibitem[{\citenamefont{Gottwald et~al.}(2004)\citenamefont{Gottwald, Likos,
  Kahl, and L\"{o}wen}}]{gottwald:prl:2004}
\bibinfo{author}{\bibfnamefont{D.}~\bibnamefont{Gottwald}},
  \bibinfo{author}{\bibfnamefont{C.~N.} \bibnamefont{Likos}},
  \bibinfo{author}{\bibfnamefont{G.}~\bibnamefont{Kahl}}, \bibnamefont{and}
  \bibinfo{author}{\bibfnamefont{H.}~\bibnamefont{L\"{o}wen}},
  \bibinfo{journal}{Phys. Rev. Lett.} \textbf{\bibinfo{volume}{92}},
  \bibinfo{pages}{068301} (\bibinfo{year}{2004}).

\bibitem[{\citenamefont{Mladek et~al.}(2008)\citenamefont{Mladek, Kahl, and
  Likos}}]{mladek:prldendrim}
\bibinfo{author}{\bibfnamefont{B.~M.} \bibnamefont{Mladek}},
  \bibinfo{author}{\bibfnamefont{G.}~\bibnamefont{Kahl}}, \bibnamefont{and}
  \bibinfo{author}{\bibfnamefont{C.~N.} \bibnamefont{Likos}},
  \bibinfo{journal}{Phys. Rev. Lett.} \textbf{\bibinfo{volume}{100}},
  \bibinfo{pages}{028301} (\bibinfo{year}{2008}).

\bibitem[{\citenamefont{Likos et~al.}(2007)\citenamefont{Likos, Mladek,
  Gottwald, and Kahl}}]{likos:jcp:2007}
\bibinfo{author}{\bibfnamefont{C.~N.} \bibnamefont{Likos}},
  \bibinfo{author}{\bibfnamefont{B.~M.} \bibnamefont{Mladek}},
  \bibinfo{author}{\bibfnamefont{D.}~\bibnamefont{Gottwald}}, \bibnamefont{and}
  \bibinfo{author}{\bibfnamefont{G.}~\bibnamefont{Kahl}},
  \bibinfo{journal}{J.~Chem.~Phys.} \textbf{\bibinfo{volume}{126}},
  \bibinfo{pages}{224502} (\bibinfo{year}{2007}).

\bibitem[{\citenamefont{Mladek et~al.}(2006)\citenamefont{Mladek, Gottwald,
  Kahl, Neumann, and Likos}}]{mladek:prl:2006}
\bibinfo{author}{\bibfnamefont{B.~M.} \bibnamefont{Mladek}},
  \bibinfo{author}{\bibfnamefont{D.}~\bibnamefont{Gottwald}},
  \bibinfo{author}{\bibfnamefont{G.}~\bibnamefont{Kahl}},
  \bibinfo{author}{\bibfnamefont{M.}~\bibnamefont{Neumann}}, \bibnamefont{and}
  \bibinfo{author}{\bibfnamefont{C.~N.} \bibnamefont{Likos}},
  \bibinfo{journal}{Phys. Rev. Lett.} \textbf{\bibinfo{volume}{96}},
  \bibinfo{pages}{045701} (\bibinfo{year}{2006}).

\bibitem[{\citenamefont{Bernabei et~al.}(2013)\citenamefont{Bernabei, Bacova,
  Moreno, Narros, and Likos}}]{bernabei:sm:2013}
\bibinfo{author}{\bibfnamefont{M.}~\bibnamefont{Bernabei}},
  \bibinfo{author}{\bibfnamefont{P.}~\bibnamefont{Bacova}},
  \bibinfo{author}{\bibfnamefont{A.~J.} \bibnamefont{Moreno}},
  \bibinfo{author}{\bibfnamefont{A.}~\bibnamefont{Narros}}, \bibnamefont{and}
  \bibinfo{author}{\bibfnamefont{C.~N.} \bibnamefont{Likos}},
  \bibinfo{journal}{Soft Matter} \textbf{\bibinfo{volume}{9}},
  \bibinfo{pages}{1287} (\bibinfo{year}{2013}).

\bibitem[{\citenamefont{Cardinaux et~al.}(2011)\citenamefont{Cardinaux,
  Zaccarelli, Stradner, Bucciarelli, Farago, Egelhaaf, Sciortino, and
  Schurtenberger}}]{cardinaux:jpcb:2011}
\bibinfo{author}{\bibfnamefont{F.}~\bibnamefont{Cardinaux}},
  \bibinfo{author}{\bibfnamefont{E.}~\bibnamefont{Zaccarelli}},
  \bibinfo{author}{\bibfnamefont{A.}~\bibnamefont{Stradner}},
  \bibinfo{author}{\bibfnamefont{S.}~\bibnamefont{Bucciarelli}},
  \bibinfo{author}{\bibfnamefont{B.}~\bibnamefont{Farago}},
  \bibinfo{author}{\bibfnamefont{S.~U.} \bibnamefont{Egelhaaf}},
  \bibinfo{author}{\bibfnamefont{F.}~\bibnamefont{Sciortino}},
  \bibnamefont{and}
  \bibinfo{author}{\bibfnamefont{P.}~\bibnamefont{Schurtenberger}},
  \bibinfo{journal}{J. Phys. Chem. B} \textbf{\bibinfo{volume}{115}},
  \bibinfo{pages}{7227} (\bibinfo{year}{2011}).

\bibitem[{\citenamefont{Coslovich et~al.}(2012)\citenamefont{Coslovich,
  Bernabei, and Moreno}}]{coslovich:jcp:2012}
\bibinfo{author}{\bibfnamefont{D.}~\bibnamefont{Coslovich}},
  \bibinfo{author}{\bibfnamefont{M.}~\bibnamefont{Bernabei}}, \bibnamefont{and}
  \bibinfo{author}{\bibfnamefont{A.~J.} \bibnamefont{Moreno}},
  \bibinfo{journal}{J. Chem. Phys.} \textbf{\bibinfo{volume}{137}},
  \bibinfo{pages}{184904} (\bibinfo{year}{2012}).

\bibitem[{\citenamefont{Briels}(2009)}]{briels:sm:2009}
\bibinfo{author}{\bibfnamefont{W.~J.} \bibnamefont{Briels}},
  \bibinfo{journal}{Soft Matter} \textbf{\bibinfo{volume}{5}},
  \bibinfo{pages}{4401} (\bibinfo{year}{2009}).

\bibitem[{\citenamefont{Kremer and Grest}(1990)}]{kremergrest:jcp:1990}
\bibinfo{author}{\bibfnamefont{K.}~\bibnamefont{Kremer}} \bibnamefont{and}
  \bibinfo{author}{\bibfnamefont{G.~S.} \bibnamefont{Grest}},
  \bibinfo{journal}{J. Chem. Phys.} \textbf{\bibinfo{volume}{92}},
  \bibinfo{pages}{5057} (\bibinfo{year}{1990}).

\bibitem[{\citenamefont{Rubinstein and Colby}(2003)}]{Rubinstein:Book2003}
\bibinfo{author}{\bibfnamefont{M.}~\bibnamefont{Rubinstein}} \bibnamefont{and}
  \bibinfo{author}{\bibfnamefont{R.~H.} \bibnamefont{Colby}},
  \emph{\bibinfo{title}{Polymer Physics}} (\bibinfo{publisher}{Oxford
  University Press, Inc}, \bibinfo{year}{2003}).

\bibitem[{\citenamefont{Hansen and {McDonald}}(2006)}]{hansen_theory_2006}
\bibinfo{author}{\bibfnamefont{J.-P.} \bibnamefont{Hansen}} \bibnamefont{and}
  \bibinfo{author}{\bibfnamefont{I.}~\bibnamefont{{McDonald}}},
  \emph{\bibinfo{title}{Theory of Simple Liquids}}
  (\bibinfo{publisher}{Academic Press}, \bibinfo{year}{2006}),
  \bibinfo{edition}{3rd} ed.

\bibitem[{\citenamefont{Hansen and Verlet}(1969)}]{hansenverlet:pr:1969}
\bibinfo{author}{\bibfnamefont{J.~P.} \bibnamefont{Hansen}} \bibnamefont{and}
  \bibinfo{author}{\bibfnamefont{L.}~\bibnamefont{Verlet}},
  \bibinfo{journal}{Phys. Rev.} \textbf{\bibinfo{volume}{184}},
  \bibinfo{pages}{151} (\bibinfo{year}{1969}).

\bibitem[{\citenamefont{Binder and Kob}(2005)}]{binder_glassy_2005}
\bibinfo{author}{\bibfnamefont{K.}~\bibnamefont{Binder}} \bibnamefont{and}
  \bibinfo{author}{\bibfnamefont{W.}~\bibnamefont{Kob}},
  \emph{\bibinfo{title}{Glassy Materials and Disordered Solids}}
  (\bibinfo{publisher}{World Scientific}, \bibinfo{year}{2005}).

\bibitem[{\citenamefont{Horbach et~al.}(2002)\citenamefont{Horbach, Kob, and
  Binder}}]{horbach:prl:2002}
\bibinfo{author}{\bibfnamefont{J.}~\bibnamefont{Horbach}},
  \bibinfo{author}{\bibfnamefont{W.}~\bibnamefont{Kob}}, \bibnamefont{and}
  \bibinfo{author}{\bibfnamefont{K.}~\bibnamefont{Binder}},
  \bibinfo{journal}{Phys. Rev. Lett.} \textbf{\bibinfo{volume}{88}},
  \bibinfo{pages}{125502} (\bibinfo{year}{2002}).

\bibitem[{\citenamefont{Moreno and Colmenero}(2006)}]{moreno_relaxation_2006}
\bibinfo{author}{\bibfnamefont{A.~J.} \bibnamefont{Moreno}} \bibnamefont{and}
  \bibinfo{author}{\bibfnamefont{J.}~\bibnamefont{Colmenero}},
  \bibinfo{journal}{J. Chem. Phys.} \textbf{\bibinfo{volume}{125}},
  \bibinfo{pages}{164507} (\bibinfo{year}{2006}).

\bibitem[{\citenamefont{Mayer et~al.}(2009)\citenamefont{Mayer, Sciortino,
  Likos, Tartaglia, L\"{o}wen, and Zaccarelli}}]{mayer:macro:2009}
\bibinfo{author}{\bibfnamefont{C.}~\bibnamefont{Mayer}},
  \bibinfo{author}{\bibfnamefont{F.}~\bibnamefont{Sciortino}},
  \bibinfo{author}{\bibfnamefont{C.~N.} \bibnamefont{Likos}},
  \bibinfo{author}{\bibfnamefont{P.}~\bibnamefont{Tartaglia}},
  \bibinfo{author}{\bibfnamefont{H.}~\bibnamefont{L\"{o}wen}},
  \bibnamefont{and}
  \bibinfo{author}{\bibfnamefont{E.}~\bibnamefont{Zaccarelli}},
  \bibinfo{journal}{Macromolecules} \textbf{\bibinfo{volume}{42}},
  \bibinfo{pages}{423} (\bibinfo{year}{2009}).

\bibitem[{\citenamefont{Coslovich et~al.}(2011)\citenamefont{Coslovich, Kahl,
  and Krakoviack}}]{crowded-jpcm}
\bibinfo{author}{\bibfnamefont{D.}~\bibnamefont{Coslovich}},
  \bibinfo{author}{\bibfnamefont{G.}~\bibnamefont{Kahl}}, \bibnamefont{and}
  \bibinfo{author}{\bibfnamefont{V.}~\bibnamefont{Krakoviack}},
  \bibinfo{journal}{J. Phys.: Condens. Matter} \textbf{\bibinfo{volume}{23}},
  \bibinfo{pages}{230302} (\bibinfo{year}{2011}).

\end{thebibliography}

\end{document}